\documentclass[aps,prx,twocolumn,superscriptaddress,notitlepage,nopacs,amsmath,amstex,amssymb,citeautoscript,longbibliography,floatfix,letter,twocolumn]{revtex4-2}

\usepackage{enumitem}
\usepackage{natbib}
\usepackage[english]{babel}
\usepackage{letltxmacro}
\usepackage{latexsym}
\LetLtxMacro{\ORIGselectlanguage}{\selectlanguage}
\makeatletter
\DeclareRobustCommand{\selectlanguage}[1]{\@ifundefined{alias@\string#1}
    {\ORIGselectlanguage{#1}}
    {\begingroup\edef\x{\endgroup
       \noexpand\ORIGselectlanguage{\@nameuse{alias@#1}}}\x}}
\newcommand{\definelanguagealias}[2]{\@namedef{alias@#1}{#2}}
\makeatother
\definelanguagealias{en}{english}
\definelanguagealias{English}{english}

\usepackage{amsthm}
\usepackage{comment}
\usepackage{graphicx}
\usepackage{bm}
\usepackage{physics}
\usepackage{braket}
\usepackage[dvipsnames]{xcolor}
\usepackage{algorithm}
\usepackage{algpseudocode}
\usepackage{blindtext}
\usepackage[normalem]{ulem}
\makeatletter
\newcommand{\printfnsymbol}[1]{\textsuperscript{\@fnsymbol{#1}}}
\makeatother
\usepackage{hyperref}
\hypersetup{
    unicode=false,
    pdftoolbar=false,
    pdfmenubar=true,
    pdffitwindow=false,
    pdfstartview={FitH},
    pdftitle={Superdiffusive Energy Transport in Kinetically Constrained Models},
    pdfauthor={Marko Ljubotina, Jean-Yves Desaules, Maksym Serbyn, Zlatko Papic},
    pdfsubject={},
    pdfcreator={},
    pdfproducer={},
    pdfkeywords={PXP, energy transport, superdiffusion},
    pdfnewwindow=true,
    colorlinks=true,
    linkcolor=teal,
    citecolor=teal,
    filecolor=teal,
    urlcolor=teal
}

\usepackage{xstring}
\newcommand{\st}[3]{{|{\StrSubstitute{#1}{u}{\uparrow}[\mystr]\StrSubstitute{\mystr}{d}{\downarrow}}\rangle_{#3}\langle{\StrSubstitute{#2}{u}{\uparrow}[\mystr]\StrSubstitute{\mystr}{d}{\downarrow}}|_{#3}}}
  
\newcommand{\id}{\text{\usefont{U}{bbold}{m}{n}1}}

\begin{document}
\title{Superdiffusive Energy Transport in Kinetically Constrained Models }

\author{Marko Ljubotina}
\thanks{M.L. and J.-Y.D. contributed equally to this work.}
\affiliation{Institute of Science and Technology Austria (ISTA), Am Campus 1, 3400 Klosterneuburg, Austria}
\author{Jean-Yves Desaules}
\thanks{M.L. and J.-Y.D. contributed equally to this work.}
\affiliation{School of Physics and Astronomy, University of Leeds, Leeds LS2 9JT, UK}
\author{Maksym Serbyn}
\affiliation{Institute of Science and Technology Austria (ISTA), Am Campus 1, 3400 Klosterneuburg, Austria}
\author{Zlatko Papi\'c}
\affiliation{School of Physics and Astronomy, University of Leeds, Leeds LS2 9JT, UK}

\date{\today}
\begin{abstract} 
Universal nonequilibrium properties of isolated quantum systems are typically probed by studying transport of conserved quantities, such as charge or spin, while transport of energy has received considerably less attention. 
Here, we study infinite-temperature energy transport in the kinetically-constrained PXP model describing Rydberg atom quantum simulators. 
Our state-of-the-art numerical simulations, including exact diagonalization and time-evolving block decimation methods, reveal the existence of two distinct transport regimes. 
At moderate times, the energy-energy correlation function displays periodic oscillations due to families of eigenstates forming different su(2) representations hidden within the spectrum. 
These families of eigenstates generalize the quantum many-body scarred states found in previous works and leave an imprint on the infinite-temperature energy transport. 
At later times, we observe a long-lived \emph{superdiffusive} transport regime that we attribute to the proximity of a nearby integrable point.
While generic strong deformations of the PXP model indeed restore diffusive transport, adding a strong chemical potential intriguingly gives rise to a well-converged superdiffusive exponent $z\approx3/2$. 
Our results suggest constrained models to be potential hosts of novel transport regimes and call for developing an analytic understanding of their energy transport.
\end{abstract}
\maketitle

\section{Introduction}

The understanding of out-of-equilibrium properties of many-body systems is one of the central problems in quantum statistical physics.
The universal aspects of nonequilibrium dynamics are commonly probed by quantum transport at infinite temperature. 
Generic chaotic models typically exhibit diffusive transport of conserved quantities such as spin~\cite{Prosen_2012, Karrasch_2014, Znidaric16, Friedman_2020, De_Nardis_2020, bertini2021finite, Wei_2022}, charge~\cite{Prosen_2012, Friedman_2020} or energy~\cite{Karrasch_2014, Lux2014, Gu_2017, Blake_2017, Friedman_2020}. 
On the other hand, disorder can give rise to slower than diffusive (subdiffusive) dynamics or even localization~\cite{Basko06, Demler14, Reichman15, Imbrie16, Znidaric16, Bar_Lev_2017, quasiperiodic2018}.

In contrast to (sub)diffusion, faster-than-diffusive transport typically rests on the existence of special structures. 
In one dimension, integrable models~\cite{sutherland2004beautiful} can support ballistic transport since their mascroscopic number of conserved quantities may prevent currents from decaying. 
Furthermore, intermediate behavior between diffusion and ballistic transport can arise in integrable models with certain symmetries, where superdiffusive Kardar-Parisi-Zhang (KPZ) dynamics has been observed~\cite{Znidaric_2011, Ljubotina_2017, Ilievski_2018, Ljubotina_2019, De_Nardis_2019, Dupont_2020, Bulchandani_2020_2, Bulchandani_2020, De_Nardis_2020, Bulchandani_2021, Wei_2022, Scheie2021, Keenan_2022}. 
Importantly, all examples of faster-than-diffusive dynamics in short range models rely on integrability. 
The same, naturally, does not hold for long-range models where superdiffusion has also been observed and explained by classical arguments using L\'evy flights~\cite{Schuckert2020,Joshi2022}.

In addition to disorder and integrability, it was recently shown that kinetic constraints may also lead to a different class of dynamics known as ``quantum many-body scars'' (QMBS)~\cite{Serbyn2021,MoudgalyaReview,ChandranReview}. 
For instance, the so-called PXP model~\cite{FendleySachdev, Lesanovsky2012} that describes constrained dynamics in Rydberg atom quantum simulators~\cite{Bernien2017, Bluvstein2021}, displays weak ergodicity breaking due to the presence of periodic revivals in the dynamics and the existence of nonthermalizing eigenstates in its spectrum~\cite{Turner2018a, wenwei18TDVPscar, Iadecola2018, Lin2019, IadecolaMagnons}. 
These rare nonergodic eigenstates were understood as forming a single approximate su(2) algebra representation, embedded into the spectrum of the Hamiltonian, even though the latter has no  SU(2) symmetry~\cite{Choi2018,Turner2018b}. 
In addition to the phenomenon of scarring, recent studies of infinite temperature charge transport in random constrained Floquet models revealed several examples of slower-than-diffusive dynamics~\cite{Singh_2021,Pal22}.

In this paper we study energy transport in the PXP model.  To access dynamics at long times, we use time-evolving block decimation (TEBD)~\cite{Vidal07, Schollwock} to calculate the energy-energy correlation function. 
Intuitively, this setup tracks the spreading of a small energy ``hump'' created at the central site of the chain, atop of the infinite temperature density matrix.
At short times the spreading of the initial energy inhomogeneity is characterized by oscillations that we attribute to the existence of multiple approximate su(2) algebra representations hidden in the spectrum of the PXP model. 
The eigenstates forming these representations encompass the previously identified QMBS  eigenstates~\cite{Turner2018a,Turner2018b}, but also include additional eigenstates corresponding to lower-spin representations. 
We relate the multiple su(2) representations to the oscillatory behavior observed in energy transport at infinite temperature, mirroring the QMBS revivals observed in quenches from special initial states~\cite{Bernien2017}. 
At later times, the oscillations due to multiple su(2) representations are damped and we observe faster-than-diffusive decay of the energy density.
The superdiffusive (possibly transient) dynamics provides further evidence to the existence of an integrable point, proximate to the PXP model~\cite{Khemani2018}. 
Surprisingly, while deforming the PXP model with a chemical potential removes the remnants of integrability, instead of restoring diffusion, it leads to a stable superdiffusive regime with a dynamical exponent $z\approx3/2$. 

The remainder of this paper is organized as follows. 
In Sec.~\ref{sec:pxp} we introduce the PXP model and present the results for its energy transport at infinite temperature. 
In Sec.~\ref{sec:shorttime} we discuss the short-time regime of the energy transport, which is characterized by oscillations that we attribute to multiple su(2) representations. 
In Sec.~\ref{sec:integrable} we analyze the late-time energy transport by applying integrable deformations to the PXP model. 
The effect of the chemical potential is studied in Sec.~\ref{sec:superdiffusion}, where it is shown to lead to a robust regime of superdiffusive transport at the accessible time scales. 
Our conclusions are presented in Sec.~\ref{sec:discussion}, while the appendices contain further details on the numerics simulations, construction of the multiple su(2) representations, and larger-range Rydberg blockade that leads to diffusive energy transport. 

\section{Infinite-temperature energy transport in the PXP model}\label{sec:pxp}

The Hamiltonian of the PXP model~\cite{Lesanovsky2012}
\begin{equation}
    H_{\rm PXP}= \Omega \sum_{i} P_{i-1}\sigma^x_iP_{i+1},
    \label{eq:PXP}
\end{equation}
operates on a chain of $N$ spins-1/2, where $P_i=\st{d}{d}{i}$ is a local projector on the $\ket{\downarrow}$ state and $\sigma^x_i$ is the corresponding Pauli matrix.  The projectors $P_i$ encode the Rydberg blockade mechanism~\cite{Labuhn2016} and lead to a block-diagonal structure of $H_\mathrm{PXP}$, also known as Hilbert space fragmentation~\cite{Sala2019,Khemani2019,Moudgalya_2021}: 
any two consecutive up spins, $|{\uparrow\uparrow}\rangle$, remain frozen under the dynamics generated by $H_{\rm PXP}$, effectively disconnecting the chain at that point. 
In what follows, we will set the Rabi frequency $\Omega=1$ and work in the \emph{reduced} Hilbert space, i.e., the largest sector that excludes any consecutive pairs of up spins. 
The PXP model in the reduced Hilbert space exhibits the repulsion of energy levels typical of chaotic systems~\cite{Turner2018a}. 

We probe energy transport via the connected energy correlation function
\begin{eqnarray}\label{eq:corr}
\langle h_0(0)h_\ell(t)\rangle_c &=& \langle h_0(0)h_\ell(t)\rangle-\langle h_0(0)\rangle\langle h_\ell(t)\rangle,
\end{eqnarray} 
where $h_\ell(0)= P_{\ell-1}\sigma^x_\ell P_{\ell+1}$ is the energy density operator at site $\ell$, and $h_\ell(t) = e^{iH_\text{PXP}t}h_\ell(0)e^{-i H_\text{PXP}t}$. 
Crucially, the expectation values in Eq.~(\ref{eq:corr}) are evaluated by taking the trace with respect to the infinite temperature density matrix within the \emph{reduced} Hilbert space. 
Specifically, we define expectation value of a given operator $O$
\begin{equation}
    \langle O \rangle \equiv \tr(\mathcal{P} O \ldots),
    \label{eq:corr-def}
\end{equation}
where the global projector $\mathcal{P} = \prod_i (\id_{i,i+1}-n_in_{i+1})$, with $n_i=|{\uparrow}\rangle_i\langle{\uparrow}|_i$, annihilates any states with two neighboring up spins. 

The projection on the reduced Hilbert space is a crucial difference with respect to the earlier studies of particle transport in constrained models, e.g., in Ref.~\cite{Singh_2021}. 
The existence of many disconnected sectors in the Hilbert space is expected to slow down the  transport. 
Indeed, below we observe diffusive or superdiffusive transport, in contrast to slower dynamics observed in Ref.~\cite{Singh_2021}. 

\begin{figure}[t]
    \centering
    \includegraphics[width=\columnwidth]{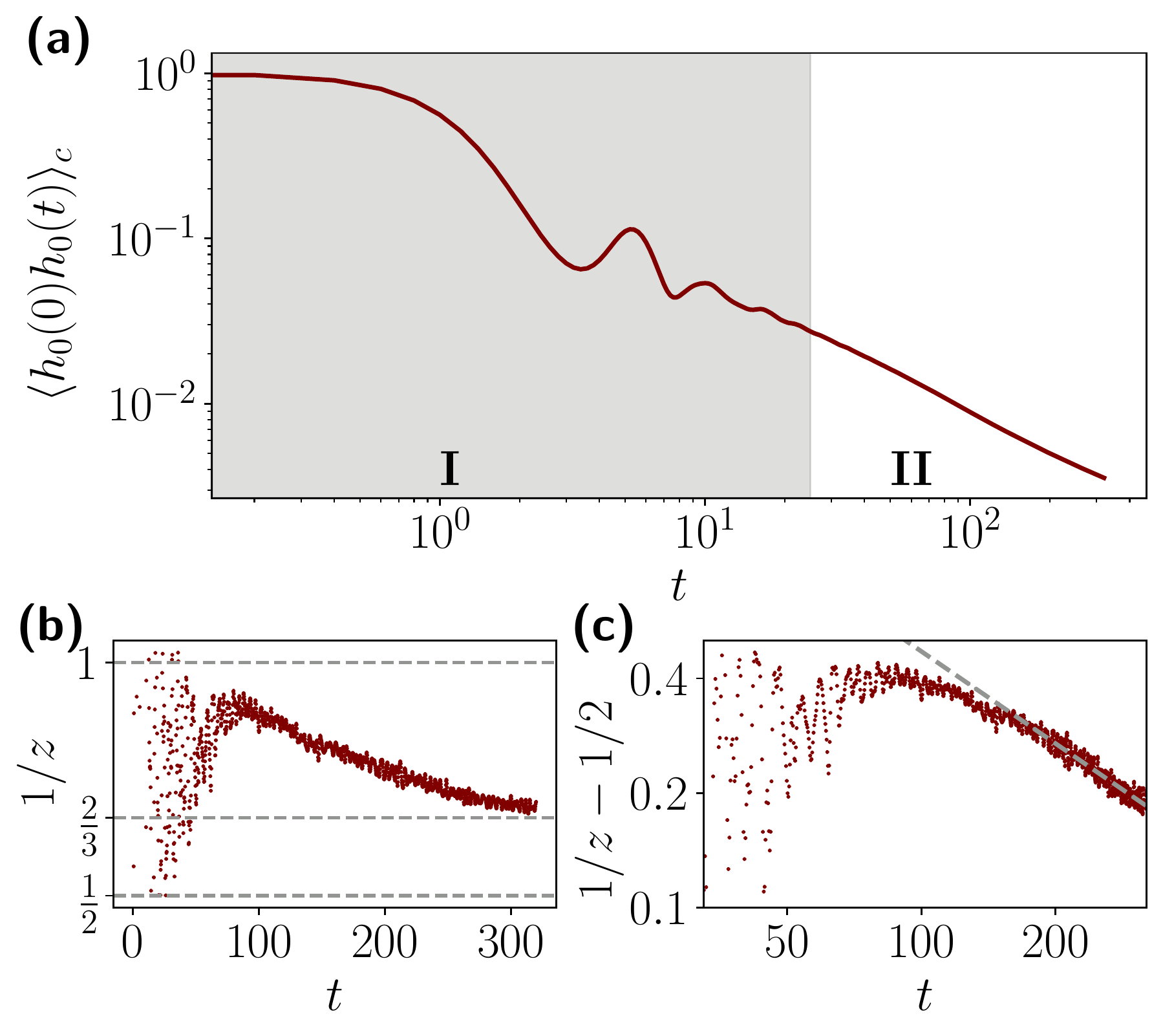}
    \caption{
        (a) Connected energy autocorrelation function in the PXP model has a short-time oscillatory regime (I) followed by power-law decay (II). 
        (b) The inverse instantaneous dynamical exponent extracted from the correlation function approaches the ballistic value of one at $t\sim100$, followed by slow decay at later times that appears to saturate to a superdiffusive value $1/z\approx 2/3$. 
        (c) Double-logarithmic plot of the data in (b) shows that power-law convergence to diffusion $1/z=1/2$ is also consistent with the data. Dashed line corresponds to $\propto t^{-0.8}$ dependence. The data is for a chain with $N=1024$ sites and bond dimension of $\chi=512$. 
    }
    \label{fig:1}
\end{figure}

Accessing energy transport in the thermodynamic limit requires the evaluation of the connected correlation function~(\ref{eq:corr}) in large systems at late times. 
To access the required system sizes and times, we use a state-of-the-art massively parallel implementation of the TEBD algorithm~\cite{Vidal07,Schollwock} based on the ITensor library~\cite{itensor}.
This allows us to simulate operator dynamics in the PXP model up to times exceeding $t \gtrsim 300$, requiring $N=1024$ lattice sites to avoid the operator spreading reaching the boundary of the system (see Appendix~\ref{app:tebd} for further details on the implementation and convergence of the data with bond dimension).

Fig.~\ref{fig:1}(a) highlights two distinct regimes in the decay of the connected energy autocorrelation function for the PXP model. 
At short times, marked by the shaded area, we observe oscillatory behavior which we explain in the following section. 
At long times, these oscillations disappear and the correlation function settles to a power-law like decay. 
This decay is conveniently probed via the instantaneous dynamical exponent 
\begin{equation}
    z^{-1}(t)=-\frac{{\rm d}\ln\langle h_0(0)h_0(t)\rangle_c}{{\rm d}\ln t},
    \label{eq:z}
\end{equation}
that gives the running exponent of the power-law decay. 
Fig.~\ref{fig:1}(b) shows that $1/z$ first approaches the ballistic value $z=1$ before relaxing slowly to a smaller value. 
Despite the decrease of $1/z$, its value remains superdiffusive ($z<2$) even at extremely long times $t\approx300$, at which the correlation function has spread approximately 300 sites from the center. 
Nevertheless, plotting the data on a log-log scale in Fig.~\ref{fig:1}(c), one cannot rule out  power-law relaxation of $z$ to diffusion at times that are inaccessible to our numerics. 

\section{Short-time oscillations and multiple su(2) representations}\label{sec:shorttime}

Unlike generic thermalizing models, the PXP model displays unusual sensitivity of its dynamics on the initial state, which has attracted considerable attention~\cite{Serbyn2021}. 
While most initial states undergo fast thermalizing dynamics in the PXP model, special states such as the N\'eel state, $|\mathbb{Z}_2\rangle = |{\uparrow}{\downarrow}{\uparrow}{\downarrow}\ldots\rangle$, feature long-lived quantum revivals~\cite{Bernien2017}, accompanied by a slow growth of entanglement~\cite{Turner2018a}. 
The non-thermalizing dynamics were explained by the existence of $N+1$ special QMBS eigenstates embedded throughout the spectrum~\cite{Turner2018a}.
However, these QMBS eigenstates constitute a vanishing fraction of the total number of states in the exponentially large Hilbert space, hence they are not expected to affect transport properties in the thermodynamic limit. 
Remarkably, the short-time oscillations in the regime I of Fig.~\ref{fig:1}(a) bear striking parallels with the scarred quantum revivals. 
In what follows we show that the PXP model hosts a much larger set of non-thermalizing eigenstates that cluster around approximately equidistant energies. 
These ``towers of states'' account for the oscillations in Fig.~\ref{fig:1}.

\begin{figure}[tb]
    \centering
    \includegraphics[width=\columnwidth]{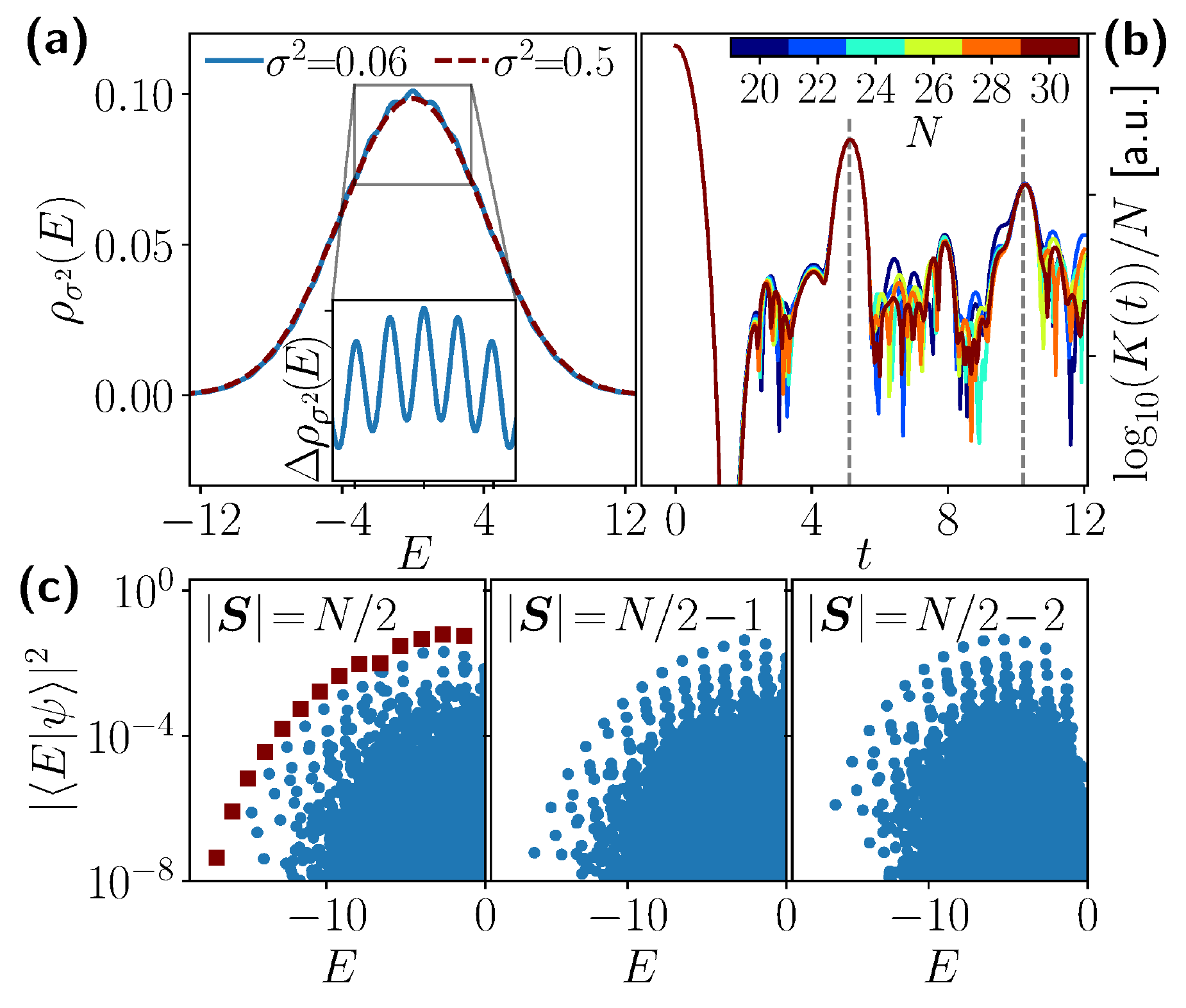}
    \caption{
        Signatures of multiple su(2) representations in the PXP model. (a) Oscillations in the smoothened density of states for $N=28$ sites, with the inset showing the difference between different sDOS (see text). 
        (b) Spectral form factor for various system sizes. The dashed lines indicate the approximate times of the peaks in Fig.~\ref{fig:1} and Fig.~\ref{fig:3} below. 
        (c) Overlap between eigenstates of the global spin-1 $S^z$ operator and the PXP eigenstates. 
        The red squares indicate the primary QMBS eigenstates with high overlap on the N\'eel state.
        All data is obtained by exact diagonalization of the PXP model with periodic boundary conditions. 
    }
    \label{fig:2}
\end{figure}

To reveal the bulk spectral properties of the PXP model, we consider a smoothened density of states (sDOS) and spectral form factor (SFF)~\cite{Mehta2004}. 
The sDOS is defined as 
\begin{equation}\label{eq:sdos}
\rho_{\sigma^2}(E) = (1/\mathcal{D}) \sum_n \exp(-(E-E_n)^2/2\sigma^2)/\sqrt{2 \pi \sigma^2},    
\end{equation}
where $E_n$ are eigenenergies, $\mathcal{D}$ is the reduced Hilbert space dimension, and $\sigma$ sets the smoothing interval. 
Fig.~\ref{fig:2}(a) shows that sDOS has tiny oscillations in the middle of the spectrum. 
These oscillations can be made more prominent by subtracting the sDOS at low and high variances,  $\Delta\rho=\rho_{\sigma^2=0.06}-\rho_{\sigma^2=0.5}$, plotted in the inset. 
The energy difference between the peaks in sDOS roughly coincides with the oscillation period of the energy correlation function in Fig.~\ref{fig:1}(a).
A similar timescale is also observed in the SFF, defined as 
\begin{eqnarray}\label{eq:sff}
K(t)=\sum_{n,m} e^{-i(E_n-E_m) t},    
\end{eqnarray}
plotted in Fig.~\ref{fig:2}(b) for a range of system sizes. 
The SFF shows clear peaks at times $t\approx 5.1$ and $t\approx 10.2$, indicated by dashed lines. 
Although the SFF, in general, is not a self-averaging quantity, the peaks appear converged in system size, and thus they are expected to persist in the thermodynamic limit. 

Oscillations in sDOS and SFF noted above can be explained by the clustering of eigenstates into towers that are approximately equally spaced in energy. 
We explain these towers of states as stemming from additional approximate su(2) representations that generalize the family of $N+1$ QMBS eigenstates identified in Ref.~\cite{Choi2018}. 
To explicitly construct multiple su(2) representations, we use the dimer picture from Ref.~\cite{Lin2019} and project the free spin-1 paramagnet with $N/2$ particles onto the constrained Hilbert space~\cite{Omiya2022}. 
The equivalent spin-1 model is obtained from PXP by mapping the Hilbert space of adjacent pairs of spin-1/2 onto that of spin-1 as $\ket{\uparrow\downarrow}{=}\ket{-}$, $\ket{\downarrow\downarrow}{=}\ket{0}$, and $\ket{\downarrow\uparrow}{=}\ket{+}$. 
The global spin operators in the spin-1 representation are defined as $S^\alpha=\sum_{b \in \Lambda_B}S^\alpha_b$, where $\Lambda_b$ is the set of non overlapping spin-1/2 pairs and $S^\alpha_b$ are spin-1 operators (see Appendix~\ref{app:su2} for details).
For the PXP Hamiltonian in Eq.~(\ref{eq:PXP}), the projection of the maximal total spin eigenstates of $S^x$ gives an excellent approximation to the original $N+1$ QMBS states of the PXP model. 
Crucially, we find that other eigenstates of $S^x$ with a large, but smaller than maximum total spin, $|\bm{S}|=|S^x|=N/2-d$, with $d\ll N$, also provide a good approximation to the PXP eigenstates after projection onto the subspace excluding $|{+}{-}\rangle$ states. 
In particular, this construction with $d=1$ explains two other sets of eigenstates identified in the PXP model in Ref.~\cite{Mondragon2021}.

The effective spin-1 model also provides simple states that can be used to identify different families of QMBS eigenstates via the enhanced overlap, see Fig.~\ref{fig:2}(c). 
The N\'eel state can be obtained as the projection of the eigenstates of spin-1 $S^z$ operator with $|\bm{S}|=|S^z|=N/2$. 
Similarly, we project an eigenstate of $S^z$ with $|\bm{S}|=|S^z|=N/2-d$ to obtain a simple state with high overlap on $N+1-2d$ eigenstates. 
Fig.~\ref{fig:2}(c) shows the overlaps of three projected eigenstates of $S^z$ for $d=0,1,2$ with the eigenstates of the PXP model.
The $d=1$ state is the superposition of all single-spin flips with momentum $k$, 
\begin{equation}
\ket{S^z{=}N/2{-}1}_k \propto \sum_{j=1}^N e^{ikj}\sigma^-_j |\mathbb{Z}_2\rangle.
\end{equation}
The anomalous overlap of these states with some eigenstates was already noted in \cite{Surace2020}, but the connection with multiple su(2) representations was not established.
For any $k\neq 0$ we obtain a state with similar properties that is the highest weight state of its own approximate su(2) representation. 
Note that states in the second and third panel of Fig.~\ref{fig:2}(c) have zero overlap with the usual scarred states as they live in different momentum sectors. 
Initial states with $d\geq 2$ have more complicated forms due to the non-trivial effect of the projection to the constrained Hilbert space. 

\section{Integrable deformations of the PXP model}\label{sec:integrable}
To further probe the relevance of multiple su(2) representations for short-time dynamics as well as the dynamical exponent governing the long-time energy transport, in this section we study two types of integrable deformations of the PXP model, which are expected to give rise to ballistic transport. 
We first consider the PNPNP deformation, which turns the PXP model into the hard-square integrable model~\cite{FendleySachdev}. 
As expected, this deformation leads to a clear ballistic exponent $z=1$ and flat spatial profile of the correlation function. 
As a second example, we study the PXPZ deformation which was numerically observed to push the energy level statistics closer to the Poisson distribution~\cite{Khemani2018}. 
This suggests the existence of a proximate integrable model, however the evidence for integrability beyond the level statistics has remained elusive. 
We note that these two integrability-enhancing deformations exert an opposite effect on the su(2) representations: the PNPNP deformation monotonically destroys such representations (and the associated QMBS revivals), while a weak PXPZ deformation leads to a strong enhancement of the su(2) representations~\cite{Choi2018}. 

\subsection{Transport at the hard-square integrable point}
\label{sec:ballistic}
\begin{figure}[b]
	\centering
	\includegraphics[width=0.95\columnwidth]{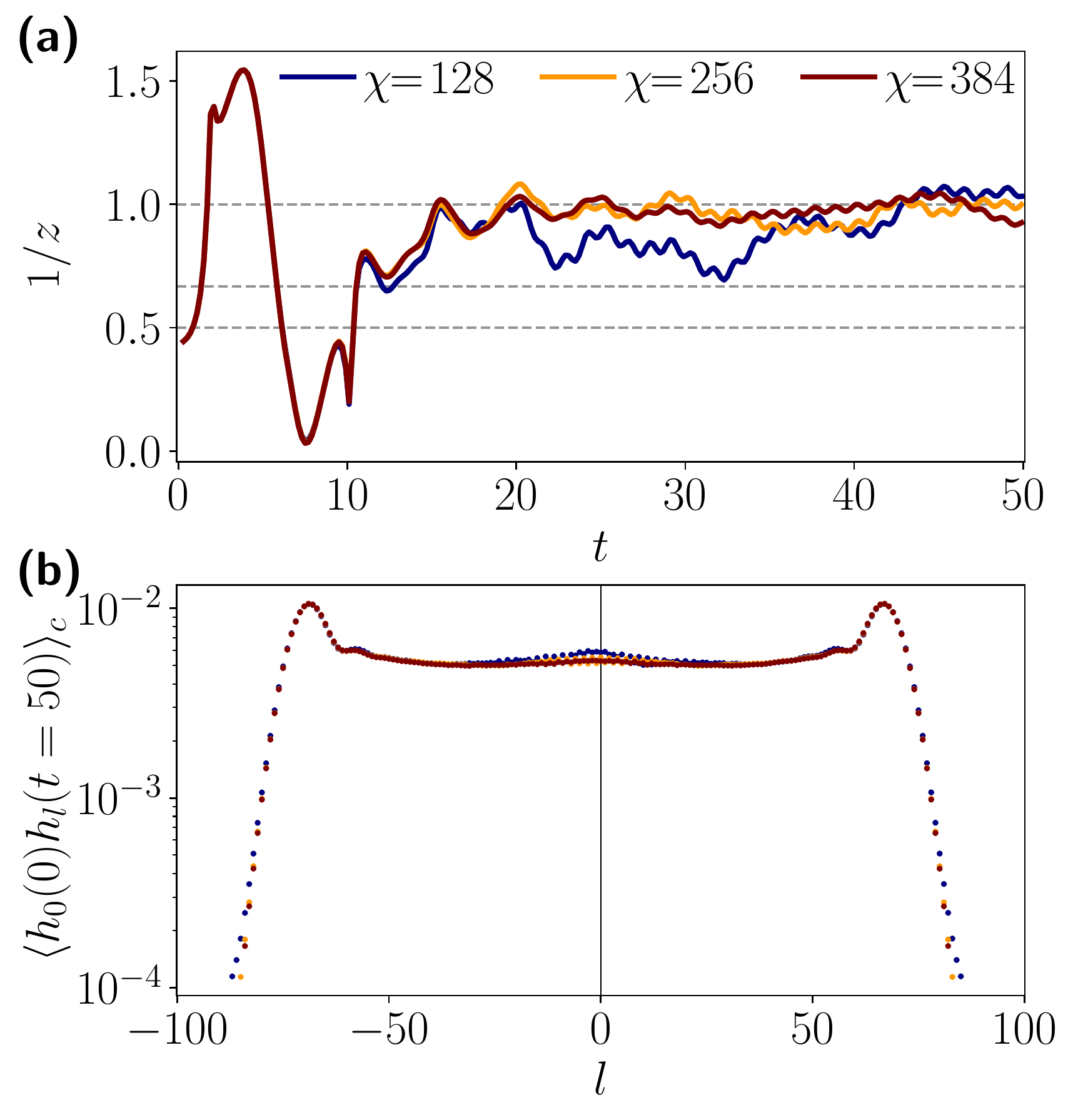}
	\caption{
		(a) Instantaneous transport exponent for the integrable PNPNP deformation of the PXP model in Eq.~(\ref{eq:pnpnp}). 
		(b) Spatial dependence of the energy autocorrelation function. 
		The profile is distinctly flat, as often seen in integrable models~\cite{Bertini_2016}.
	}
	\label{fig:c2}
\end{figure}

In order to benchmark our study of transport we consider the integrable hard-square model~\cite{FendleySachdev}, that can be obtained as a deformation of the PXP model
\begin{equation}\label{eq:pnpnp}
	H_\mathrm{PNPNP} = H_{\rm PXP}+\xi\sum_iP_{i-2}n_{i-1}P_{i}n_{i+1}P_{i+2},
\end{equation}
for the particular value of the deformation strength $\xi=\pm1$. 
In Fig.~\ref{fig:c2}(a) we show that the integrable point $\xi=1$ is characterized by a stable ballistic exponent $z=1$ after an initial transient. 
Note that before settling down to the value $z=1$,  there is an initial overshoot to values $z(t)>1$, well-converged across different bond dimensions. 
At later times, however, the higher bond dimensions of $\chi=256$ and 384 yield nearly identical results, while $\chi=128$ data is different, suggesting that this value is insufficient for capturing the dynamics.

\begin{figure*}[tb]
	\centering
	\includegraphics[width=\textwidth]{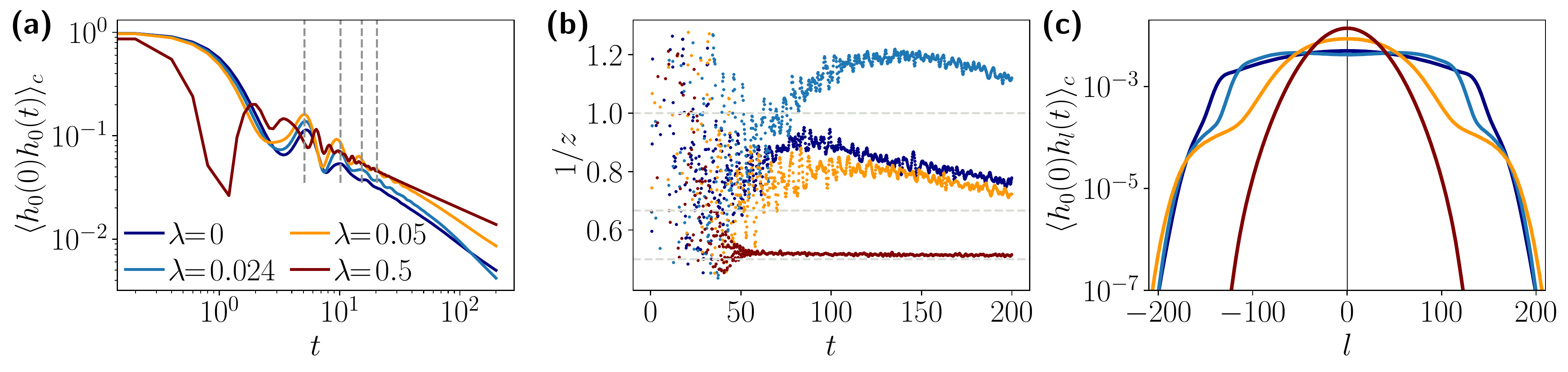}
	\caption{
		(a) Connected energy auto-correlation function for several strengths $\lambda$ of the PXPZ deformation. Early time oscillations approximately peak at times $t\in\{5.1,10.2,15.3,20.4\}$ (dashed lines) for the su(2)-enhancing perturbation $\lambda=0.05$. 
		(b) Long time decay is fastest for $\lambda=0.024$, as manifested by the inverse dynamical exponent approaching the value of one from above. 
		In contrast, large $\lambda=0.5$ results in a rapid onset of diffusive dynamics. 	
		(c) Spatial dependence of the connected energy correlation function at $t=200$ in the PXP model with the PXPZ deformation. 
		In the vicinity of $\lambda\approx0.026$, the profile becomes visibly flat, which is typical of ballistic transport. 
        The data is for $N=768$ and bond dimensions $\chi=384$ [panels (a), (b)] and $\chi=256$ in (c). 
	}
	\label{fig:3}
\end{figure*}

The spatial profile in Fig.~\ref{fig:c2}(b) provides additional support for integrability. 
The profiles are converged for the two largest  bond dimensions and show weak spatial dependence, which is often seen in integrable models such as the XXZ model~\cite{Bertini_2016}.
Having benchmarked our approach on a known integrable model, we next study a different deformation that leads to a similar transport phenomenology but does not correspond to an exact integrable model.

\subsection{PXPZ deformation}

To further probe the relevance of multiple su(2) representations, we deform the model with a local PXPZ perturbation,
\begin{equation}
    H_\mathrm{PXPZ}=H_\mathrm{PXP}-\lambda \sum_i \left(\sigma^z_{i-2}+\sigma^z_{i+2}\right) P_{i-1}\sigma^x_iP_{i+1}.
    \label{eq:PXPZ}
\end{equation}
Previously, it was shown that $\lambda\approx0.05$ stabilizes the highest-spin su(2) representation~\cite{Choi2018}. 
In Appendix~\ref{app:su2}, we demonstrate that other representations with a lower total spin are also stabilized by a comparable value of $\lambda$. 
Remarkably, Ref.~\cite{Khemani2018} observed that value of $\lambda \approx 0.024$ leads to the onset of Poisson level statistics, conjecturing the existence of a nearby integrable model. 
Naturally, both the enhancement of su(2) structure and the presence of a nearby integrable point are expected to leave an imprint on the energy transport.

Fig.~\ref{fig:3}(a) illustrates the effect of the PXPZ deformation on the energy autocorrelation function and the dynamical exponent. 
We observe that the deformation strength $\lambda=0.05$ gives the strongest enhancement of oscillations in the early time regime, further confirming that oscillations are caused by multiple su(2) representations. 
However, the long-time value of $1/z$ for $\lambda=0$ and $0.05$ behave nearly identically. 
In contrast, the value $\lambda=0.024$ only weakly enhances the oscillations but yields a much faster decay of the correlation function at late times. 
The extracted $1/z$ exponent in Fig.~\ref{fig:3}(b) overshoots the ballistic value $z=1$ and converges to it from above, consistent with a proximate integrable point. 
Finally, a large deformation $\lambda=0.5$ leads to fast saturation of $1/z\approx 0.5$ corresponding to conventional diffusion. 

In the vicinity of an integrable point, superdiffusion 
can appear as a crossover between nearly-ballistic behavior at short times and diffusion at late times~\cite{Friedman_2020, Ferreira2020, Durnin2021} (see also the review~\cite{Bastianello2021}). 
The intuition is that transport should at first behave as in a corresponding integrable model until the system starts to feel the effect of the integrability-breaking perturbation, which leads to slow quasiparticle decay processes.
To corroborate this picture of ballistic transport followed by slow decay, we computed the spatial profiles of the correlation function in the PXP model with the PXPZ deformation in Fig.~\ref{fig:3}(c). 
As seen in this figure, the profile exhibits nearly flat dependence on position, similar to what we observed in the integrable hard square model. 

\section{Stable superdiffusion}\label{sec:superdiffusion}

The proximity of an integrable point naturally explains the observed long timescales in the dynamics of the PXP model. 
In order to avoid slow convergence towards the thermodynamic limit, we must consider stronger deformations of the model. 
While a large PXPZ deformation restores diffusion in Fig.~\ref{fig:3}, such a perturbation is not readily available in experiments. 
Instead, we focus on the PNP deformation
\begin{equation}
    H_\mathrm{PNP}=H_{\rm PXP}+\delta\sum_iP_{i-1}n_iP_{i+1}.
    \label{eq:PNP}
\end{equation}
The PNP term counts the number of Rydberg excitations in the constrained Hilbert space, hence this term represents the chemical potential. 
The latter has been used to probe the dynamics of the PXP model in Rydberg atom arrays~\cite{Bluvstein2021} and Bose-Hubbard optical lattices~\cite{Su2022}. 
We note that the deformation in Eq.~(\ref{eq:PNP}) cannot be expressed using the generators of the approximate su(2) structure~\cite{Choi2018}. 
As such, it destroys the peaks in the sDOS already for small values of $\delta$ and it has been shown to make the dynamics from the $\ket{\mathbb{Z}_2}$ state ergodic~\cite{Yao2022}.

\begin{figure}[tb]
	\centering
	\includegraphics[width=\columnwidth]{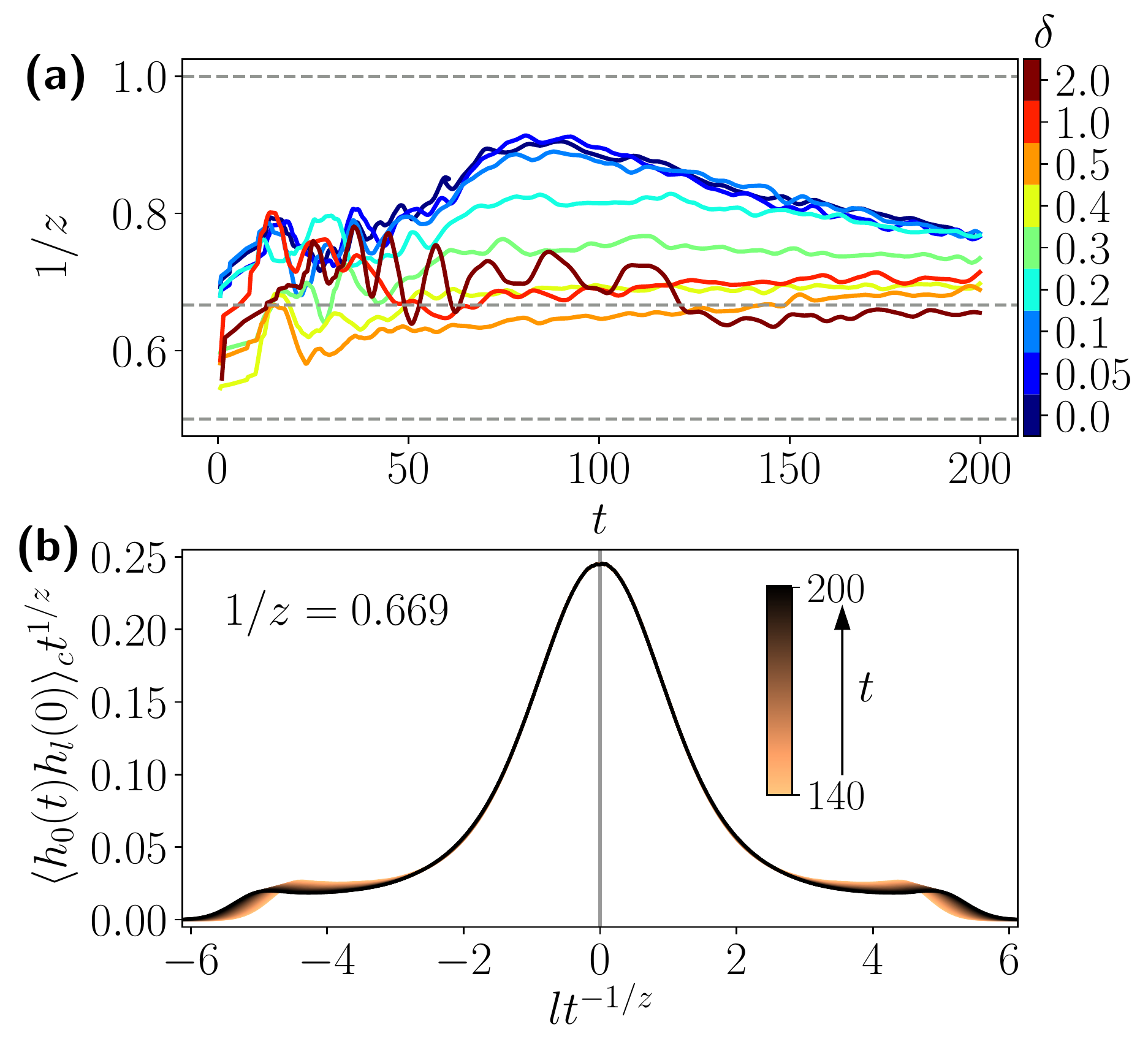}
	\caption{
		Stable superdiffusive energy transport for a large PNP deformation of the PXP model. 
		(a) Sufficiently large deformations $\delta\ge0.4$ lead to a clear superdiffusive exponent $z\approx1.5$. 
		(b) Single-parameter scaling of the spatial profiles of the connected correlation function for $\delta=0.5$ shows a collapse for $1/z\approx0.669$ (average value of $1/z$ in the interval $t\in[100,200]$). 
		The ballistically propagating peaks at large $|x|$ are expected to disappear as $t\to\infty$. 
		The data is for $N=1024$ sites with bond dimensions $\chi=384$ for $\delta=2$ and $\chi=256$ for $\delta\leq 1$. 
		For clarity purposes, data in panel (a) has been smoothed using a Gaussian filter. 
	}
	\label{fig:4}
\end{figure}
Fig.~\ref{fig:4}(a) shows the instantaneous dynamical exponent for a range of deformation parameters $\delta\in[0,2]$. 
For weak deformations $\delta\leq 0.2$, the exponent varies slowly, similar to the PXP model in Fig.~\ref{fig:1}(b). 
However, the effect of a nearby integrable point diminishes once the deformation is sufficiently strong. 
Surprisingly, for large deformations $\delta\ge0.4$, we observe clear superdiffusive transport with a well-converged dynamical exponent $z\approx1.5$. 
The robust superdiffusion, observed over a broad range of PNP perturbations, suggests that even the PXP model itself may have $z\approx1.5$, although much longer times may be needed to observe the convergence of the exponent to this value. 
In Appendix~\ref{app:deformations} we explore several other deformations of the model that similarly give rise to a persistent superdiffusive regime at all accessible timescales for weak deformations. 

The transport exponent $z=3/2$ is one of the hallmarks of the KPZ universality class, which has recently been observed in integrable quantum systems with certain symmetries~\cite{Znidaric_2011, Ljubotina_2017, Ljubotina_2019, Dupont_2020, Wei_2022}. 
To further test if $H_\mathrm{PNP}$ belongs to the same universality class, we analyze the spatial profile of the correlation function in Fig.~\ref{fig:4}(b). 
The profiles of correlation function at different times show a clear collapse with a single scaling parameter that matches the average value of $1/z$ in the interval $t\in[100,200]$. 
However, the presence of ballistic peaks visible at the edges of numerically computed profiles in Fig.~\ref{fig:4}(b) prohibits us from reliably discriminating between the theoretically expected KPZ or Gaussian scaling functions. 

Finally, we check the hypothesis that the unusual transport may originate from the Hamiltonian~(\ref{eq:PNP}) being the projection of a free paramagnet (with both transverse and longitudinal fields) onto the reduced Hilbert space. 
To this end, in Appendix~\ref{app:ppxpp} we consider a similar model projected onto the reduced Hilbert space corresponding to a range-2 Rydberg blockade. 
The constraint now excludes both nearest-neighbor as well as next-nearest-neigbor up spins, leading to an effective PPXPP model with a chemical potential. 
Surprisingly, the deformed PPXPP model reveals clear diffusion for a broad range of chemical potential values, as demonstrated in Appendix~\ref{app:ppxpp}, suggesting that a simple projection of a free paramagnetic Hamiltonian does not guarantee superdiffusive transport.

\section{Discussion}\label{sec:discussion}
	
We studied energy transport in the largest connected component of the Hilbert space of the PXP model and its deformations. 
We explained the observed oscillations in the short-time dynamics via towers of eigenstates that form multiple approximate su(2) representations. 
These towers enlarge the set of known QMBS states in the PXP model and lead to observable signatures of scarring in infinite-temperature transport. 
The long-time behavior is shown to be affected by a nearby integrable point, confirming that the PXPZ deformation gives rise to nearly ballistic transport, as suggested by the level statistics indicators~\cite{Khemani2018}. 

Strong deformations with the chemical potential were shown to move the PXP model away from an integrable point but, surprisingly, gave rise to a broad regime of superdiffusive energy transport. 
The observed transport exponent, $z\approx3/2$,  is tantalizingly close to the value corresponding to KPZ universality class. 
Nevertheless, it remains unclear why the PXP model would belong to this class, given the existing examples of KPZ dynamics in integrable spin chains with SU(2) or higher symmetry. 
Furthermore, it is known that higher order corrections can lead to a transient KPZ-like exponent~\cite{Lux2014}. 
However, with perturbations such as PNP, the observed superdiffusion remains stable up to long times of $\sim 200$ in natural units $1/\Omega$, casting doubt on the relevance of higher-order corrections.
Moreover, it is not clear why the chemical potential would particularly enhance such corrections.

Finally, the link between scarring and superdiffusion remains to be understood.
On the one hand, as we explained above, the chemical potential $\delta$ destroys the su(2) structure responsible for scarring from the $|\mathbb{Z}_2\rangle$ state, while at the same time this does not appear to affect the late-time dynamical exponent, suggesting the two phenomena to be unrelated. 
On the other hand, upon further increase of $\delta$, it was observed that scarring signatures reappear but in the polarized initial state, $\ket{\downarrow\downarrow\downarrow\ldots}$~\cite{Su2022}. 
It is therefore possible that other su(2) representations -- linked to scarring from the polarized state -- emerge at larger $\delta$, suggestive of a closer intertwining between scarring and superdiffusion.

Our work opens a number of new directions related to the interplay of weak ergodicity breaking phenomena and transport in constrained models. 
In the first direction, the existence of a much larger number of nonergodic towers of  eigenstates calls for a development of their systematic theory. 
It remains to be understood if complete towers can be stabilized by the same weak deformation of the PXP model, akin to Ref.~\cite{Choi2018}, or if only individual representations can be stabilized. 
Our work suggests that several representations with large total spin are enhanced by it, but PXPZ alone is not enough to make them exact. 
If there exists a unique deformation of the PXP model that stabilizes all states in the towers, it may lead to the coexistence of integrability and su(2) algebra. 
In a more practical direction, the tower states are characterized by a larger amount of entanglement and it would be interesting to explore their applications in information-storage or quantum-enhanced metrology~\cite{Dooley2021,DesaulesMultipartite,Dooley2022}. 
Finally, it would be important to understand the underlying mechanisms for the emergence of towers in other models with non-exact scars, such as higher-spin PXP~\cite{wenwei18TDVPscar} and clock models~\cite{Bull2019}, lattice gauge theories~\cite{Surace2020,Desaules2022a,Desaules2022b}, and fractional quantum Hall states on stretched cylinders~\cite{Moudgalya2019}.
While in this paper we have implicitly focused on lattice models, we note that several recent works have explored realizations of scars in quantum field theories~\cite{Dodelson2022,Cotler2022, Delacretaz2022}, which could offer a fruitful setting for studying the relation between scarring and transport.

Concerning transport, our observations challenge the current understanding of chaotic quantum models, which are expected to exhibit diffusive (or slower) transport dynamics. 
In particular, our work suggests that certain classes of constrained models, when studied in the reduced Hilbert space, may provide stable examples of superdiffusive transport. 
At present, the explanation for the robust superdiffusive transport observed here is missing, highlighting the need for the development of a theoretical description of transport in systems with constraints.
One potential explanation for superdiffusion in chaotic models could stem from nonlinear fluctuating hydrodynamics~\cite{Spohn_2014}, provided additional conserved charges exist within the reduced Hilbert space. 
Indeed, KPZ transport has been shown to arise in a broad class of low-dimensional classical models with particle, momentum and energy conservation~\cite{vanBeijeren2012}. 
Our brute-force numerical search did not yield a clear signature of additional conservation laws in the PXP model, hence it remains unclear whether its transport fits the framework of Refs.~\cite{Spohn_2014, vanBeijeren2012}.
Another potential explanation for the unusual transport could be related to the semiclassical aspects of the PXP dynamics when the latter is projected onto the variational manifold of matrix product states~\cite{wenwei18TDVPscar}. 
The study of quantum dynamics with translationally-invariant initial conditions in Ref.~\cite{Michailidis2019} revealed the existence of large Kolmogorov-Arnold-Moser (KAM) tori in the classical phase space resulting from the variational projection. 
Provided these tori survive in the absence of translation invariance, they may play a role in energy transport. 
Finally, the current capabilities of Rydberg quantum simulators~\cite{bluvstein2022quantum} may allow to probe our predictions experimentally and to gain further insights into the energy transport, in particular in higher dimensions. 

\begin{acknowledgments}
We would like to thank Alexios Michailidis, Sarang Gopalakrishnan and Achilleas Lazarides for useful comments. 
M.L. and M.S.\ acknowledge support by the European Research Council (ERC) under the European Union's Horizon 2020 research and innovation program (Grant Agreement No.~850899).
J.-Y.D. and Z.P. acknowledge support by EPSRC grant EP/R513258/1 and the Leverhulme Trust Research Leadership Award RL-2019-015.  Statement of compliance with EPSRC policy framework on research data: This publication is theoretical
work that does not require supporting research data.
M.S., M.L. and Z.P. acknowledge support by the Erwin Schr\"odinger International Institute for Mathematics and Physics (ESI).
M.L. and M.S. acknowledge PRACE for awarding us access to Joliot-Curie at GENCI@CEA, France, where the TEBD simulations were performed. 
The TEBD simulations were performed using the ITensor library~\cite{itensor}. 
\end{acknowledgments}

\appendix

\section{Details of TEBD simulations}
\label{app:tebd}

\subsection{Implementation details}

TEBD simulations were used to directly evolve the energy density operator within the reduced space 
\begin{equation}
    h_\ell(0)=\mathcal{P}P_{\ell-1}\sigma^x_\ell P_{\ell+1}.
\end{equation}
In order to do this, we must first bring the above operator into a matrix product operator (MPO) form. 
This is a relatively trivial step since $\mathcal{P}$ can be described as a simple bond dimension 2 MPO
\begin{equation}
    \mathcal{P}_{x_0,N}=\mathcal{L}_{x_0}
    \left(\prod_{i=x_0+1}^{x_0+N-2}\mathcal{M}_i
    \right)\mathcal{R}_{x_0+N-1},
\end{equation}
where $N$ is the length of the system.
Here we defined the following matrices (used to conveniently represent tensors in MPO form since their elements are operators) 
\begin{eqnarray}
    \mathcal{L}_i=\begin{pmatrix}\st{d}{d}{i}+\st{u}{u}{i}&\st{d}{d}{i}\end{pmatrix},\\
    \mathcal{M}_i=\begin{pmatrix}\st{d}{d}{i}&\st{d}{d}{i}\\\st{u}{u}{i}&0\end{pmatrix},\\
    \mathcal{R}_i=\begin{pmatrix}\st{d}{d}{i}\\\st{u}{u}{i}\end{pmatrix}.
\end{eqnarray}
Using this notation we write the MPO for the energy density
\begin{equation}
    h(\ell,0)=\mathcal{P}_{1,\ell-2}\mathcal{D}_{\ell-1}\mathcal{X}_\ell\mathcal{D}_{\ell+1}\mathcal{P}_{\ell+2,N-\ell-1},
\end{equation}
where we additionally introduced the tensors
\begin{eqnarray}
    \mathcal{D}_i=\begin{pmatrix}\st{d}{d}{i}\end{pmatrix},\\
    \mathcal{X}_i=\begin{pmatrix}\st{d}{u}{i}+\st{u}{d}{i}\end{pmatrix}.
\end{eqnarray}
We note that a similar representation of the energy density may be easily derived for the deformations of the PXP model considered in this work. 

Notice that the PXP model has a local Hamiltonian density acting on 3 sites, as a result one would have to use 3-site propagation, requiring 2 singular value decompositions (SVD) for each site as well as multiple layers for a fourth order Trotter decomposition. 
Instead, in this work we merge pairs of sites, thus obtaining a 2-site Hamiltonian. 
Firstly, the constraint reduces the local Hilbert space dimension of two sites from 4 to 3, thus speeding up our calculations, since the bulk of the time is spent performing singular value decompositions (SVD) which scale as $\mathcal{O}(d^6\chi^3)$, where $d$ is the local Hilbert space dimension and $\chi$ is the bond dimension. 
In terms of MPOs presented above, the merging of pairs of sites is implemented by multiplying the tensors for the pairs of sites to be merged. 
Secondly, since we are using propagation of MPOs, the effective Hilbert space dimension of each pair of sites is equal to 9. 
That means that, at later times when the bond dimension saturates, we can take advantage of the faster randomized rank-reduced singular value algorithm (RRSVD)~\cite{Tamascelli_2015, Halko_2011}, allowing us to significantly speed up the simulations. 
Finally, in the case of longer range deformations, we merge more sites together to obtain an effective two-site Hamiltonian density. 

\subsection{Convergence of numerical data}

\begin{figure}[ht]
    \centering
    \includegraphics[width=0.95\columnwidth]{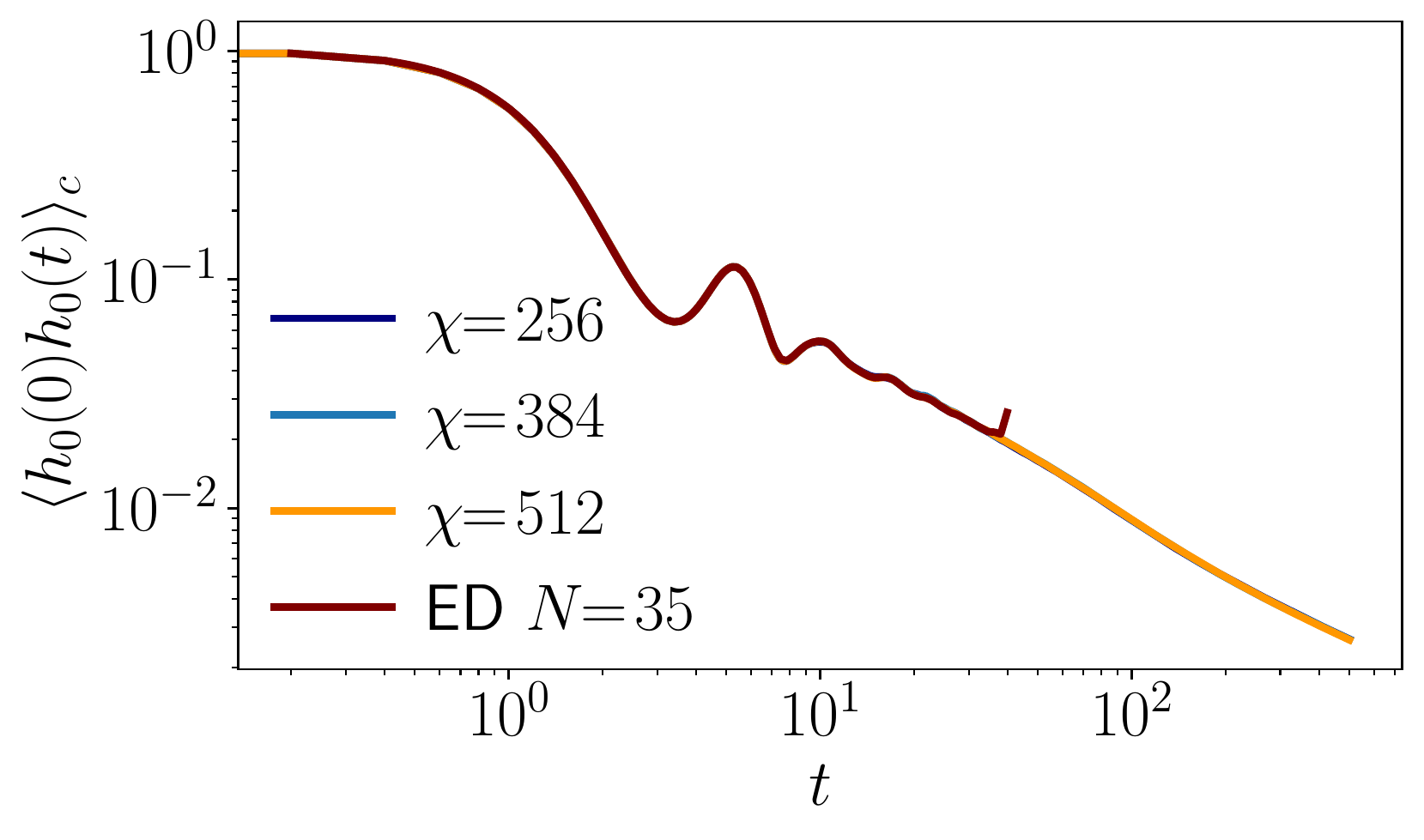}
    \caption{Comparing the PXP model results for different bond dimensions $\chi\in\{256,384,512\}$ against ED result for system size $N=35$. 
    We observe that all bond dimensions give nearly identical results with relative errors between consecutive bond dimensions at most $1\%$. }
    \label{fig:a1}
\end{figure}

Throughout this work, we use the 4th order Trotter decomposition in order to achieve high accuracy with a relatively large time step, typically taken to be $\delta t=0.2$. 
We always compute the results with several different bond dimensions and verify that the data is well-converged. 
All results are furthermore benchmarked against large-scale exact diagonalization (ED) simulations on systems with sizes up to $N=35$.
To demonstrate convergence in the bond dimension,  Fig.~\ref{fig:a1} shows the results for the PXP model for several different bond dimensions $\chi\in\{256,384,512\}$, as well as the largest available ED data. 

\section{Exact diagonalization}\label{sec:typical}

\begin{figure}[htbp]
	\centering
	\includegraphics[width=\linewidth]{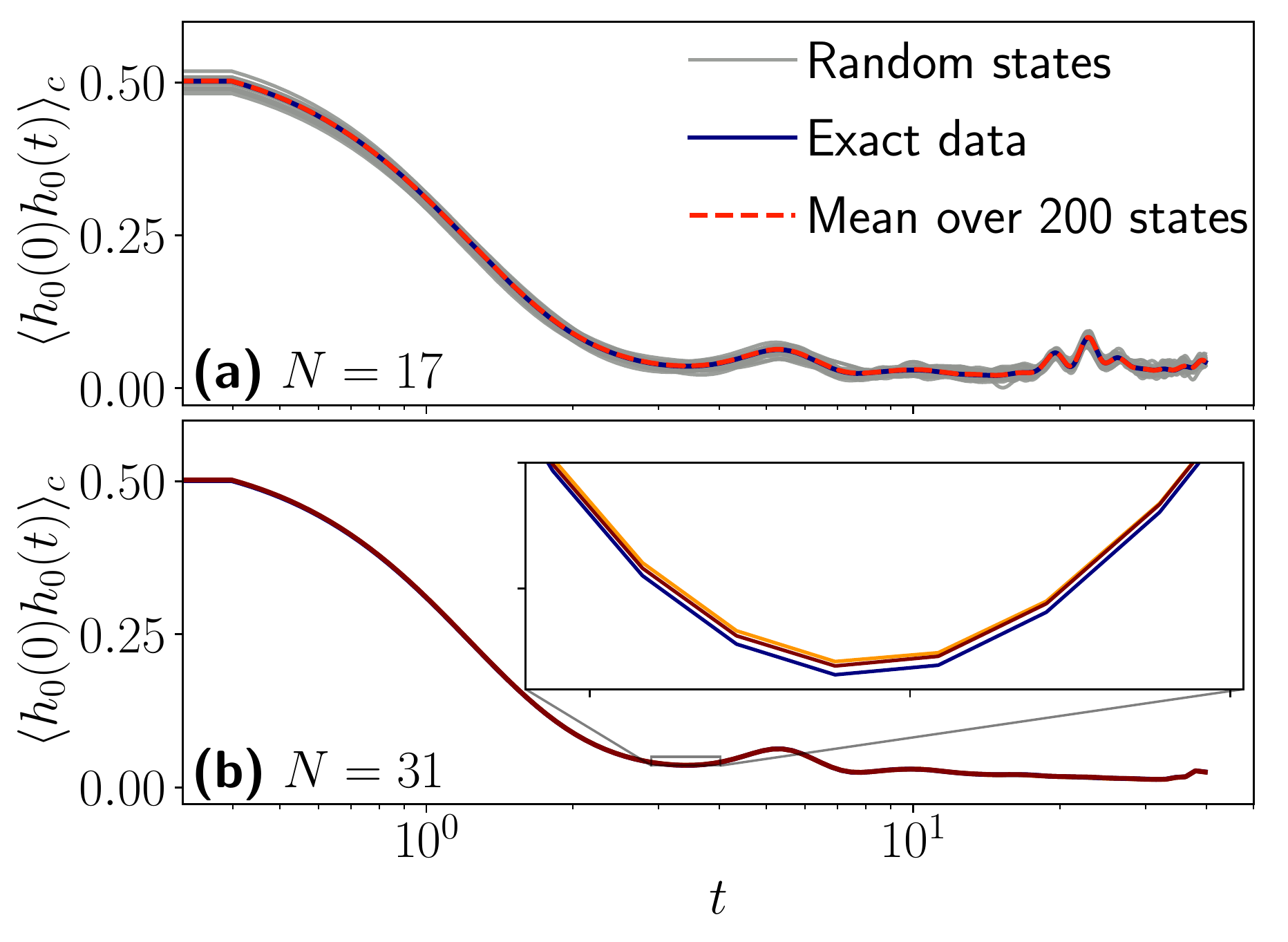}
	\caption{\small (a) Comparison of the energy correlation function between random typical states and the exact computation performed by evolving the full evolution operator. The average of random typical states matches well with the latter. (b) Energy correlation function for 3 random typical states in a large system. The inset is a zoom on one time interval, showing that different typical states give very similar results.}
	\label{fig:PXP_corr}
\end{figure}

While ED is unbiased and does not suffer from precision issues, evolving the full energy operator is computationally costly. 
To circumvent this, we use typical pure states with the same energy density profile~\cite{bartsch2009}.
To study the infinite-temperature ensemble we follow the procedure outlined in Ref.~\cite{Steinigeweg2017}. 
We define the initial state as
\begin{equation}\label{eq:psi0}
\ket{\psi}=\frac{1}{\mathcal{N}}\sum_{k=1}^\mathcal{D}c_k \ket{\phi_k},
\end{equation}
where $\ket{\phi_k}$ are the Fock basis states (of the \emph{constrained} Hilbert space), $c_k$ are complex coefficients for which the real and imaginary parts are drawn from Gaussian distributions with mean 0, and $\mathcal{N}$ is the normalization factor.  
We then compute the state $\ket{\psi^\prime}=h_l\ket{\psi}$.
By evolving both $\ket{\psi}$ and $\ket{\psi^\prime}$, we can compute the auto-correlation function at infinite temperature as 
\begin{equation}
\begin{aligned}
\braket{\psi^\prime(t)|h_r|\psi(t)}&=\braket{\psi|h_lU^\dagger (t) h_r U(t)|\psi}\\
&=\braket{\psi|h_l h_r(t)|\psi}\\
&\approx \Tr \left(h_l h_r(t) \right),
\end{aligned}
\end{equation}
where $U(t)=\exp(-iHt)$ is the time-evolution operator.
In Fig.~\ref{fig:PXP_corr} we see that the results obtained using such typical states agree very closely those obtained using the full evolution operator. 
Furthermore, in large systems all random typical states give very close results and only a few of them need to be evolved in order to  approximate the average.

\section{Multiple su(2) representations}
\label{app:su2}

Quantum scarred eigenstates in the PXP model, which are responsible for the revivals of the N\'eel initial state, have been understood from an su(2) algebra point of view in Refs.~\cite{Choi2018,Bull2020}. 
More recently, an alternative explanation based on a spin-1 parent Hamiltonian was proposed in Ref.~\cite{Omiya2022}. 
Here we generalize the latter approach in order to construct multiple su(2) representations, thereby allowing us to classify a larger number of scarred eigenstates in the PXP model that give rise to oscillations in the infinite-temperature energy transport.

Due to the Rydberg constraint, a pair of neighboring atoms can never be in the configuration $\ket{\uparrow \uparrow}$, and so the state of a dimer can be mapped to spin-1 according to  
\begin{eqnarray}
\ket{\uparrow\downarrow}{\equiv}\ket{-}, \;\; \ket{\downarrow\downarrow}{\equiv}\ket{0}, \;\; \mathrm{and} \;\; \ket{\downarrow\uparrow}{\equiv}\ket{+}.     
\end{eqnarray}
The constraint must furthermore prevent neighboring pairs $\ket{+-}=\ket{\downarrow\uparrow\uparrow\downarrow}$, so it can be written in terms of a projector
\begin{eqnarray}
 \mathcal{P}=\prod_{b \in \Lambda_B}\left(1-\ket{+-}\bra{+-}_{b,b+1}\right),   
\end{eqnarray}
where $\Lambda_B$ denotes the set of all dimers. The Hamiltonian of the parent model is then 
\begin{eqnarray}
        \notag H_{\Lambda_B}&=& \sqrt{2}\sum_{b \in \Lambda_B}S^x_b -\sum_{b \in \Lambda_B}\left(\ket{+,0}+\ket{0,-}\right)\bra{+,-}_{b,b+1} \\
       \notag  &-& \sum_{b \in \Lambda_B}\ket{+,-}\left(\bra{+,0}+\bra{0,-}\right)_{b,b+1}\\
        &\equiv& H_Z+H_1+H_2,
    \label{eq:dimer}
\end{eqnarray}
where $H_Z$ is simply the free spin-1 paramagnet, and $H_1$ and $H_2$ are 
added to cancel the matrix elements violating  the Rydberg constraint. As a consequence, $\left[\mathcal{P},H_{\Lambda_B}\right]=0$ and $H_{\Lambda_B}$  
is equivalent to the PXP model in the constrained sector. So we can define 
\begin{equation}
    \tilde{H}_\mathrm{PXP}=H_\mathrm{PXP}\oplus 0_{\perp}=\mathcal{P}H_{\Lambda_B}
\end{equation}
that acts as PXP in its sector but annihilates any state outside of it. One can also notice that $\mathcal{P}H_2=0$ but $\mathcal{P}H_1\neq 0$. 

 \begin{figure}[t]
    \centering
    \includegraphics[width=0.95\columnwidth]{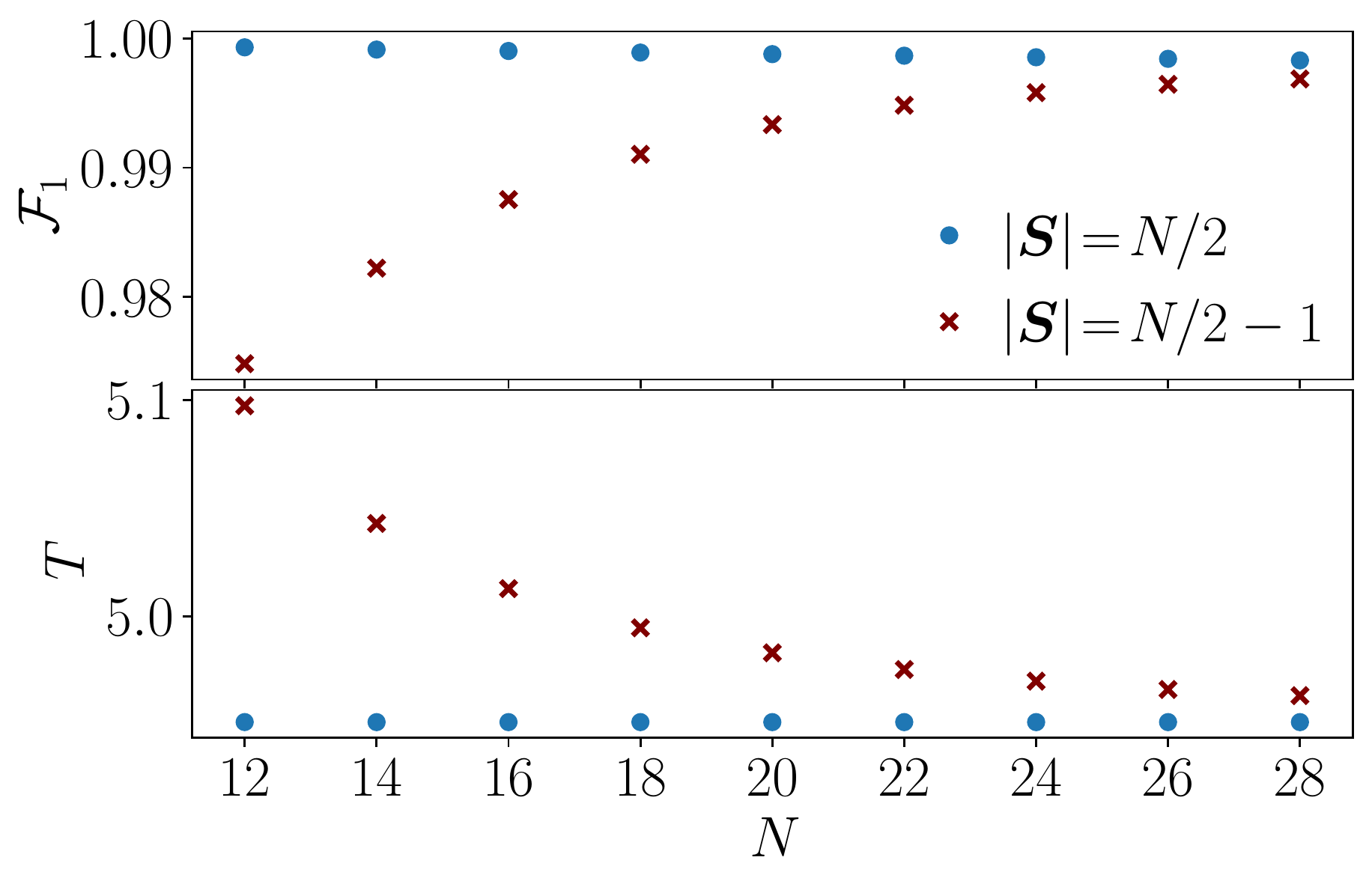}
    \caption{Fidelity and period of revivals for two eigenstates of $S^z$ projected into the PXP Hilbert space and evolved with $H_\mathrm{PXPZ}$ with $\lambda=0.051$. For $|{\bm S}|=N/2-1$ the revivals get better as system size is increased, and should theoretically match up with those of $|{\bm S}|=N/2$ in the thermodynamic limit.}
    \label{fig:su2_pert}
\end{figure}

Let us consider an eigenstate $\ket{E}$ of $H_Z$ with an energy $E$.
If there is a perturbation $\delta H$ such that $\mathcal{P} \left(H_1+\delta H\right)\ket{E}=0$ and $\left[\mathcal{P},\delta H\right]=0$, it follows that $\ket{\tilde{E}}=\mathcal{P}\ket{E}$ is an eigenstate of $\tilde{H}_\mathrm{PXP}+\delta\tilde{H}$ with energy $E$, where $\delta\tilde{H}=\mathcal{P}\delta H$. Indeed,
\begin{eqnarray}
      \notag  \left(\tilde{H}_\mathrm{PXP}{+}\delta\tilde{H} \right)\ket{\tilde{E}}&=& \left(\tilde{H}_\mathrm{PXP} +\delta\tilde{H}\right)\left(\mathcal{P}\ket{E}\right) \\
     \notag    &=& \mathcal{P}\left( H_{\Lambda_B}+\delta H \right)\ket{E} \\
    \notag    &=& \mathcal{P}\left( H_Z+H_1+H_2+\delta H \right)\ket{E} \\
     \notag   &=& \mathcal{P}H_Z\ket{E}{+}\mathcal{P}\left( H_1+H_2+\delta H \right)\ket{E} \\
        &=& \mathcal{P}H_Z\ket{E}=\mathcal{P} E\ket{E}=E\ket{\tilde{E}}.
    \label{eq:eigfun}
\end{eqnarray}
The desired perturbation 
\begin{eqnarray}\label{eq:nonhermpert}
 \delta H=\sum_{\Lambda_B}\frac{1}{2}\left(\ket{+,0}+\ket{0,-} \right)\bra{0,0}_{b,b+1}   
\end{eqnarray}
was found in Ref.~\cite{Omiya2022} for the set of scarred eigenstates that have maximal total spin.  Indeed, this perturbation holds for states in which every pair of neighboring dimers forms a spin-2 quintuplet. This is only the case if the total spin is maximal, i.e., equal to $N/2$.

\begin{figure}[t]
    \centering
    \includegraphics[width=0.95\columnwidth]{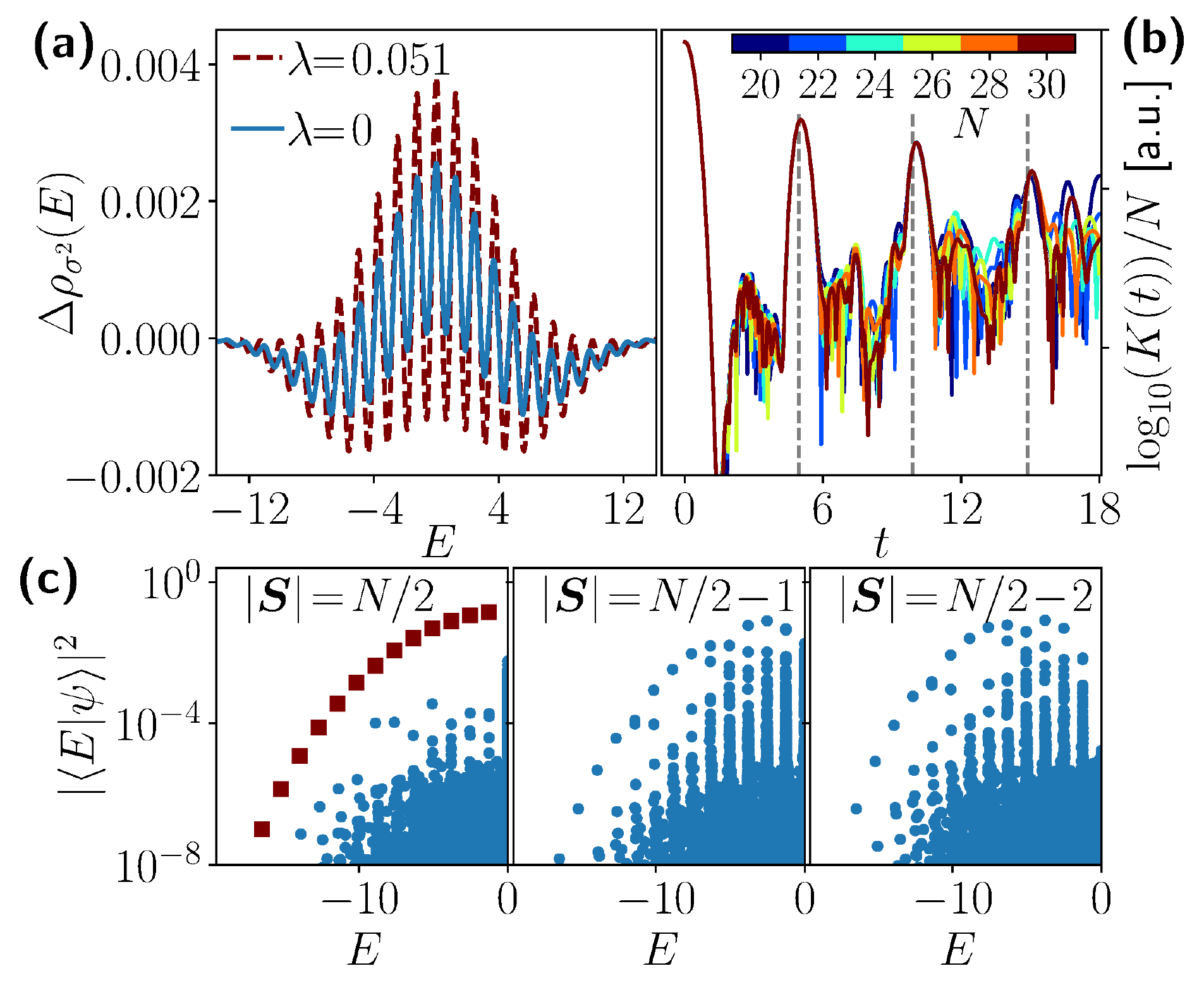}
    \caption{The effects of PXPZ perturbation on the multiple su(2) representations. (a) Difference between sDOS with and without the perturbation. 
    (b) Peaks in the SFF that are converged for various system sizes. The peaks occur at roughly the same times as the quench revivals from the N\'eel state, indicated by the dashed lines. (c) Overlap between eigenstates of $S^z$ and the eigenstates of the perturbed PXP model. The red squares indicate the usual scarred eigenstates with high overlap with the N\'eel state.}
    \label{fig:su2_fig2_pert}
\end{figure}

However the proportion of spin-2 pairs (in contrast to spin 0 or spin 1) is close to 1 for all large spin representation $|{\bm S}|=N/2-d$, $d\ll N/2$ in large systems. In particular, for a fixed value of $d$, the fraction of spin-2 pairs increases with system size. In Fig.~\ref{fig:su2_pert}, we show the fidelity at the first QMBS revival with the perturbation in Eq.~(\ref{eq:PXPZ}) with $\lambda=0.051$. For the N\'eel state with $|{\bm S}|=N/2$, the fidelity decreases as $N$ gets larger due to the Hilbert space size increasing. This is the usual behavior in systems with non-exact QMBSs. However, for $|{\bm S}|=N/2-1$, the revivals actually get \emph{better} with an increase in system size, at least in the range of $N$ probed. We understand this as stemming from the increasing fraction of spin-1 pairs forming spin-2 quintuplets.

This perturbation does not only affect a few su(2) representations near $|{\bm S}|=N/2$, but has a clear effect on the entire spectrum. Fig.~\ref{fig:su2_fig2_pert} shows similar quantities as Fig~\ref{fig:2} in the main text, but for $H_\mathrm{PXPZ}$ with $\lambda=0.051$. Panel (a) shows that the perturbation makes the towers of states much more prominent, and panel (b) shows that the oscillatory features in the spectral form factor are cleaner and also persist over longer times. It should also be noted that in the perturbed case, the period of ``revivals" in the SFF matches more closely with the revivals in a quench starting from the N\'eel state. Finally, panel (c) shows that as for the N\'eel state, the perturbation helps separate the top band of eigenstates form the more thermal bulk.
 
\begin{figure}[b]
    \centering    \includegraphics[width=0.95\columnwidth]{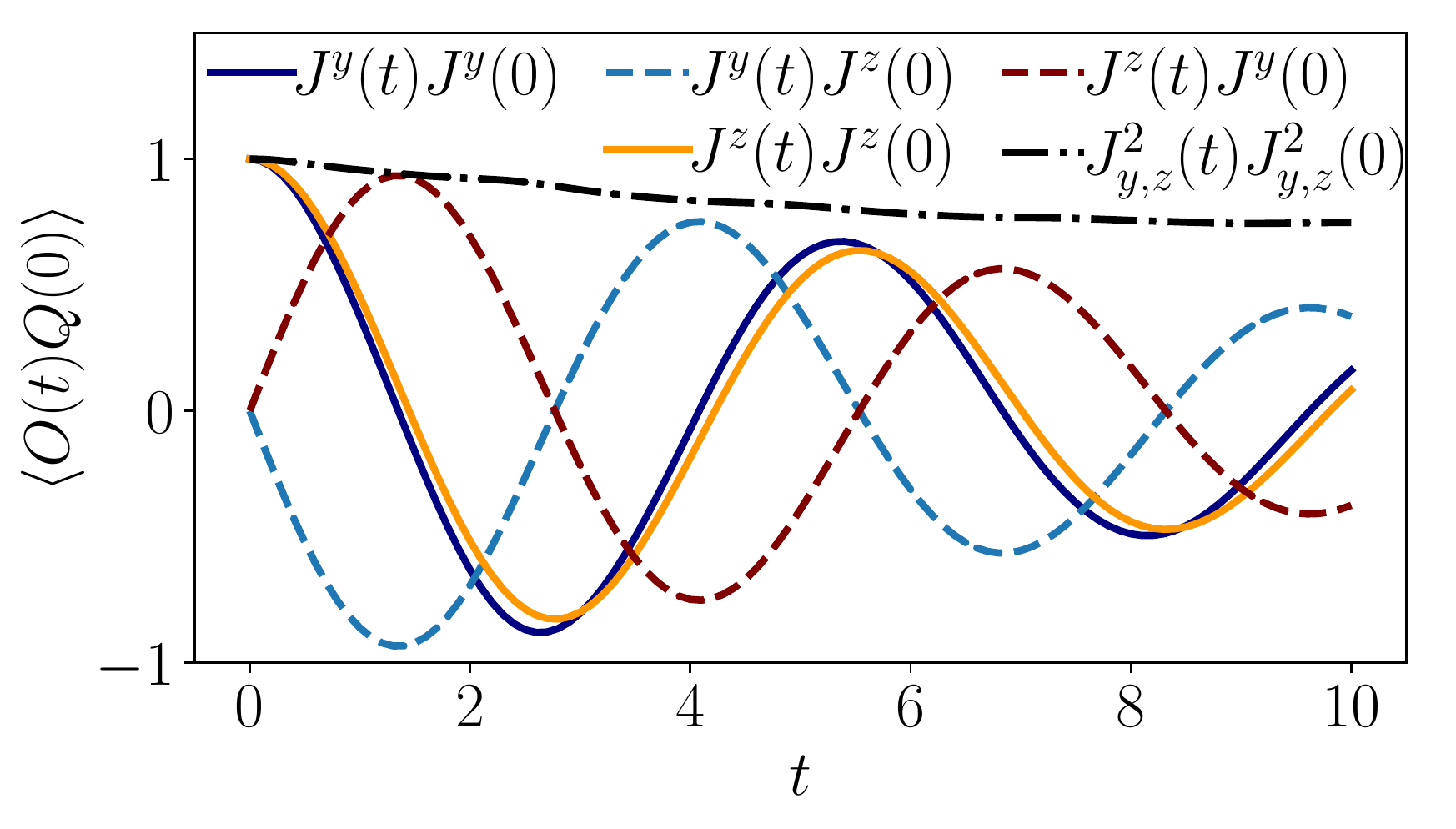}
    \caption{Time evolution of the various global spin operators projected to the constrained space. In particular, $\langle J^2_{y,z}(t) J^2_{y,z}(0) \rangle$ remains close to 1, indicating an approximate conservation law.  }
    \label{fig:su2_ops}
\end{figure}

In summary, our results show that there are multiple approximate su(2) representations in the PXP model that stem from the same spin-1 parent Hamiltonian. Another way to see that these representation leave a strong imprint in the spectrum is to look at the behavior of observables. Indeed, in the parent Hamiltonian the operators $S^x$, $S^y$ and $S^z$ obey the usual su(2) commutation relations. If sufficiently many su(2) representations survive in the PXP model, we expect the projection of these operators to at least approximately obey the same commutation relations as well. Projecting $S^y$ and $S^z$ gives the following operators
\begin{eqnarray}
    \mathcal{P}S^y\mathcal{P}& \rightarrow&\frac{1}{\sqrt{2}}\sum_j (-1)^jP_{j-1}\sigma^y_jP_{j+1}=\frac{1}{\sqrt{2}}J^y, \label{eq:Jy} \\
    \mathcal{P}S^z\mathcal{P}&\rightarrow&\frac{1}{2}\sum_j (-1)^j\sigma^z_j=J^z, \label{eq:Jz}    
\end{eqnarray}
in the PXP sector, where the dimerization gives rise to the staggering. These results match with operators that were previously devised only for the highest-spin representation in~\cite{IadecolaMagnons, Mondragon2021}, as the operators $Y_\pi / \tilde{\sigma}^y_\pi$ and $Z_\pi / \tilde{\sigma}^z_\pi$ defined in these works match with $J^y$ and $J^z$.

Now one can study how the operators in Eqs.~(\ref{eq:Jy})-(\ref{eq:Jz}) evolve in time. If the projection of the dimerized operators approximately preserves their commutation relations, we expect that
\begin{eqnarray}
 \langle J^y(t) J^y(0) \rangle\approx\langle J^z(t) J^z(0) \rangle\approx \cos(\omega t),   
\end{eqnarray}
where $\langle O Q \rangle={\rm Tr}(OQ)$ is the expectation value at infinite temperature in the constrained Hilbert space. As the operator $\left(S^y\right)^2+\left(S^z\right)^2=\bm{S}^2-\left(S^x\right)^2$ commutes with $S^x$, one can also probe if the operator $J^2_{y,z}=\left(J^y\right)^2+\left(J^z\right)^2$ gives rise to $\langle J^2_{y,z}(t) J^2_{y,z}(0) \rangle \approx 1$ as expected for a constant of motion. Fig.~\ref{fig:su2_ops} shows that all these relations are approximately obeyed in a finite system with 20 sites. 
This once again shows clearly that there are multiple su(2) representations stemming from the spin-1 parent model, as the usual N\'eel scarred states could not produce such a strong effect at infinite temperature.

\section{Other deformations leading to superdiffusion}
\label{app:deformations}

\begin{figure}[t]
    \centering
    \includegraphics[width=\columnwidth]{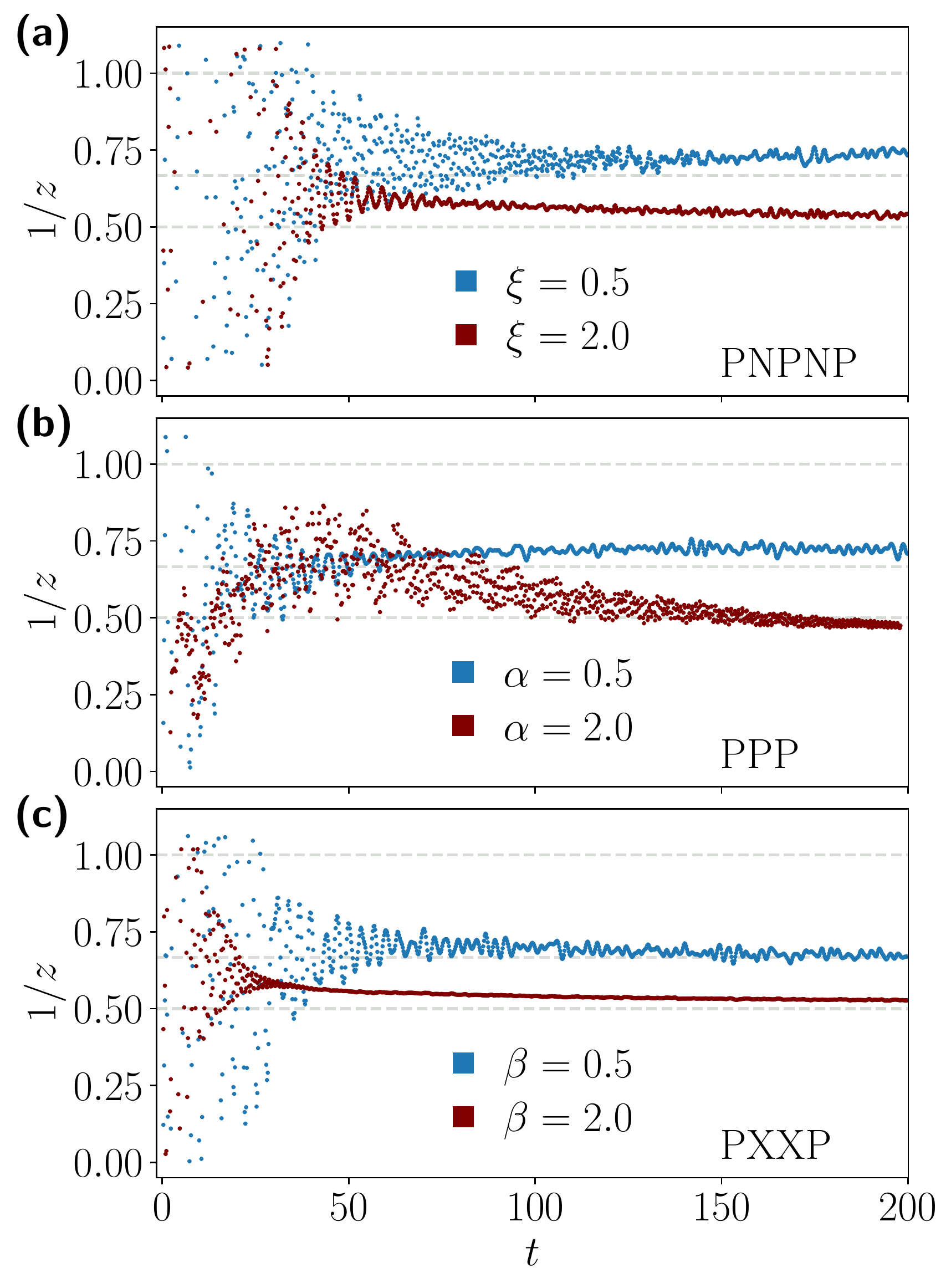}
    \caption{Instantaneous transport exponent for PNPNP, PPP and PXXP [panels (a), (b) and (c), respectively] at deformation parameters $0.5$ and $2.0$. In all cases, we observe superdiffusive transport across all accessible timescales for sufficiently weak  deformation, while stronger deformations push the dynamical exponent towards the diffusive value. Data for PNPNP and PPP were obtained with bond dimension $\chi=256$ and data for PXXP were obtained with bond dimension $\chi=192$. }
    \label{fig:d1}
\end{figure}
Interestingly, many deformations of the PXP model lead to superdiffusive dynamics at weaker values of the deformation parameter, but then typically approach diffusion as the deformation strength is increased.  This is illustrated in Fig.~\ref{fig:d1}(a) for the PNPNP deformation defined in Eq.~(\ref{eq:pnpnp}) above,  which shows 
a seemingly stable superdiffusive behavior up to long times at $\xi=0.5$ and nearly diffusive dynamics at $\xi=2$. A similar trend is seen in panels (b) and (c) of Fig.~\ref{fig:d1}, which correspond to the PPP deformation
\begin{equation}
    H_\mathrm{PPP}  =H_{\rm PXP}+\alpha\sum_iP_{i-1}P_{i}P_{i+1},
\end{equation}
and the PXXP deformation
\newpage
\begin{equation}
    H_\mathrm{PXXP} =H_{\rm PXP}+\beta\sum_iP_{i-1}\sigma^x_{i}\sigma^x_{i+1}P_{i+2},
\end{equation}
respectively. 

\section{Stronger Rydberg blockade -- the PPXPP model}
\label{app:ppxpp}

\begin{figure}[b]
	\centering
	\includegraphics[width=\columnwidth]{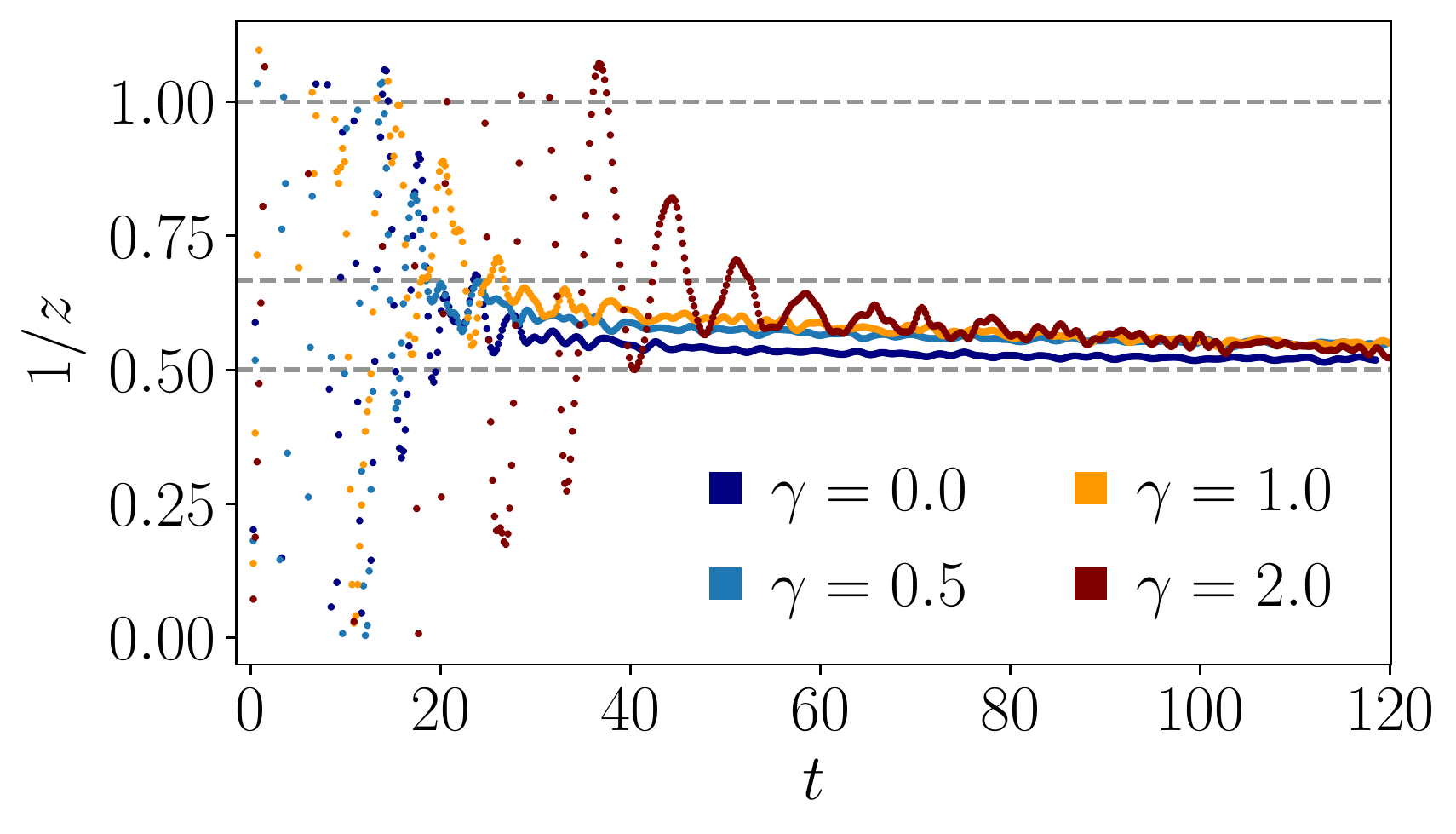}
	\caption{Instantaneous dynamical exponent in the PPXPP model with various strengths of the PPNPP deformation. We observe clearly different behavior to the PXP+PNP model with energy diffusion being observed for all values of the PPNPP deformation parameter. Data obtained with bond dimensions $\chi=256$. }
	\label{fig:e1}
\end{figure}

Finally, we present data on the PPXPP model with a PPNPP deformation
\begin{eqnarray}
	\notag    H_\mathrm{PPXPP} &=& \sum_i P_{i-2}P_{i-1}\sigma^x_{i}P_{i+1}P_{i+2} \\
	&+&\gamma\sum_iP_{i-2}P_{i-1}n_{i}P_{i+1}P_{i+2},    
\end{eqnarray}
which can be viewed as a longer-range generalization of the PXP model with a PNP deformation that gave rise to stable superdiffusive behavior. 
Interestingly, the results shown in Fig.~\ref{fig:e1} indicate that the increased range of the projectors immediately makes the model diffusive. This suggests that a more systematic study is needed to understand the relation between the existence of superdiffusive transport and the nature of the constraint present in the model. 

\bibliography{biblio_reduced}

\end{document}